\documentclass[a4paper,fleqn]{article}

\usepackage{times,cite}
\usepackage[T1]{fontenc}
\usepackage[latin9]{inputenc}
\usepackage{amsmath}
\usepackage{graphicx}
\usepackage[english]{babel}

\begin{document}

\title{On the multipole moments of a rigidly rotating 
       fluid body}

\author{Robert Filter and Andreas Kleinw\"achter \\
        Theoretisch-Physikalisches Institut, \\
        University of Jena, Max-Wien-Platz 1,
        07743 Jena, Germany}
\date{February 10, 2009}

\maketitle

\begin{abstract}
Based on numerical simulations and analytical calculations 
we formulate a new conjecture
concerning the multipole moments of a rigidly rotating fluid body in equilibrium.
The conjecture implies that the exterior region of such a fluid is not described by the Kerr metric.
\end{abstract}

\section{Introduction}

Bardeen and Wagoner \cite{BW71} observed in numerical tests in 1971
that the quadrupole moment of a rigidly rotating disk of dust is
always greater than that of the Kerr metric with the same angular
momentum and mass except in the extreme relativistic case, when they
become the same,
\begin{eqnarray}
\left|Q_{2}^{\text{B\&W}}\right| & \geq & \left|Q_{2,\text{Kerr}}^{\text{B\&W}}\right|.
\end{eqnarray}

As already concluded by Bardeen and Wagoner, the general expectation
is that this also holds for other spacetimes containing a central
rigidly rotating perfect fluid. The recent work of Bradley and Fodor \cite{BF08} lends support for this in the slowly rotating
case.

It is now interesting
to ask the question whether this holds not only for the quadrupole moment
but for every moment,
\begin{eqnarray}
\left|Q_{n}\right| & \geq & \left|Q_{n,\text{Kerr}}\right|.\label{eq:Q-Equation}
\end{eqnarray}
In this paper, based upon \cite{Filter08}, we will formulate 
this conjecture and collect evidence in favor of it
from numerical and analytical results.
We will make use of geometrical units with $G=c=1$.

\section{Formulation of the conjecture}

The multipole moments calculated by Fodor et al. in \cite{FHP89}
are equivalent to the invariantly defined ones by Geroch and Hansen
\cite{Geroch70,Hansen73} for axially symmetric and stationary spacetimes.
Because of the form of the Ernst potential on the axis, the mass-
and rotation moments are in this context given by
\begin{equation}
Q_{n,\text{Kerr}} \,=\, \text{i}^{n}\frac{J^{n}}{M^{n-1}}
 \,\equiv\, M_{n}+\text{i}J_{n}.
\end{equation}
In accordance with relation (\ref{eq:Q-Equation}) we state:

{\bf Generalized Quadrupole-Conjecture.}
{\it
For axially symmetric, stationary and asymptotically
flat spacetimes with angular momentum $J$, mass $M$ and multipole moment $Q_{n}$ 
\begin{eqnarray}
A_{n}:=\left|\frac{J^{n}}{M^{n-1}\cdot Q_{n}}\right| 
        & \leq & 1\label{eq:Gen.Eq.}
\end{eqnarray}
always holds if the spacetime is that of a rigidly rotating perfect fluid body in equilibrium, surrounded by vacuum.
}

Furthermore, in accordance with our experience,
the equality is only reached in the
case of a black hole limit, which is then necessarily an extreme
Kerr black hole, see \cite{Meinel06}. In general, the exterior spacetime differs from the Kerr metric.

\section{Evidence}

\subsection{Newtonian Limit}

In the Newtonian limit, a rigidly rotating object has angular momentum
\begin{equation}
J \,=\, \Theta\cdot\Omega
\end{equation}
with moment of inertia $\Theta$ and angular velocity $\Omega$.
In this limit we can compute a kinetic energy and define a characteristic
velocity via
\begin{equation}
2E_{\text{kin}} \,=\, \Theta\cdot\Omega^{2}
   \,=:\, M\cdot v_{\text{char}}^{2}.
\end{equation}
Restricting ourselves to axial and equatorial symmetry, relation (\ref{eq:Gen.Eq.})
becomes for $n=2,\ 4,\ \dots$
\begin{equation}
\left|\frac{J^{n}}{M^{n-1}\cdot Q_{n}}\right| \,=\, \left|\frac{\Theta^{n}\Omega^{n}}{M^{n-1}\cdot Q_{n}}\right|
 \,=\, \left|\frac{\Theta^{n/2}}{M^{n/2-1}\cdot 
       Q_{n}}\right|v_{\text{char}}^{n} 
 \,\leq\, 1,
\end{equation}
which should always hold under the condition of small velocities $v_{\text{char}}\ll1$
provided $\Theta^{n/2}M^{1-n/2}Q_{n}^{-1}$ is limited.

\subsection{Numerical tests}

With help of the numerical program described in \cite{AKM02,AKM03b} we were able to
test the conjecture for different equations of state and different
topologies in the case of the quadrupole and octupole moments. Exemplarily
we will show some of the results to underline the conjecture.

The tests covered three equations of state for spheroidal stars containing
homogeneous matter, an MIT-Bag model equation of state for quark matter following
\cite{Chodosetal74} and a model for a completely degenerated ideal
neutron gas following \cite{Stoner32} and additionally a quadrupole-test
for thin rotating rings.

\begin{figure}
\begin{center}
\includegraphics[scale=0.6]{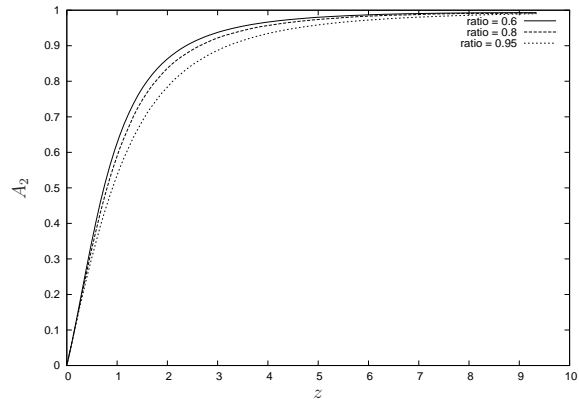}
\caption{\label{fig:rings:quadrupole}Rotating homogeneous rings   for different ratios of radii depending on the redshift
  parameter $z$, see \cite{Filter08}.}
\end{center}
\end{figure}

\newpage
Figures \ref{fig:MIT-Bag,quadrupole} and \ref{fig:rings:quadrupole}
show $A_{2}$ for the cases of a star with an MIT-Bag model equation of state and homogeneous rings, respectively.
\begin{figure}
\begin{center}
\includegraphics[scale=0.6]{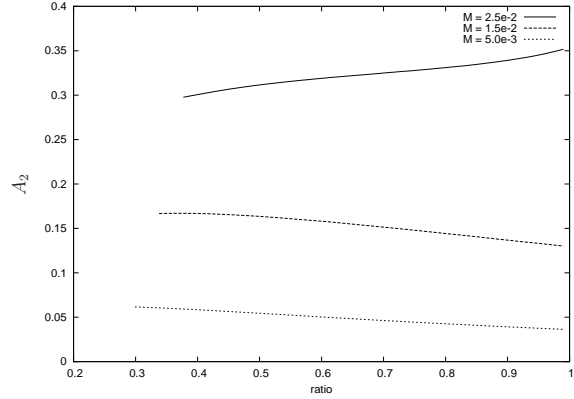}
\caption{\label{fig:MIT-Bag,quadrupole}$A_{2}$ 
          for quadrupole moments of
          strange stars with an MIT-Bag model equation of state 
          and varying masses depending on the ratio of polar 
          to equatorial radius.}
\end{center}
\end{figure}

\begin{figure}
\begin{center}
\includegraphics[scale=0.6]{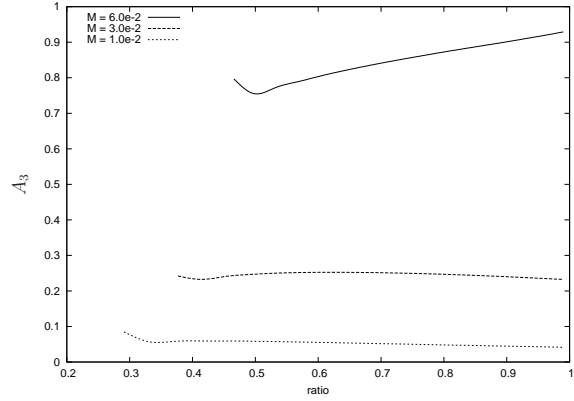}
\caption{\label{fig:octupole:homogeneous}$A_{3}$ for  
  homogeneous stars with different masses.}
\end{center}
\end{figure}

Figure \ref{fig:octupole:homogeneous} shows a test in the case of the octupole moment for homogeneous stars with different masses.

\subsection{The rigidly rotating disk of dust}

\begin{figure}
\begin{center}
\includegraphics[scale=0.7]{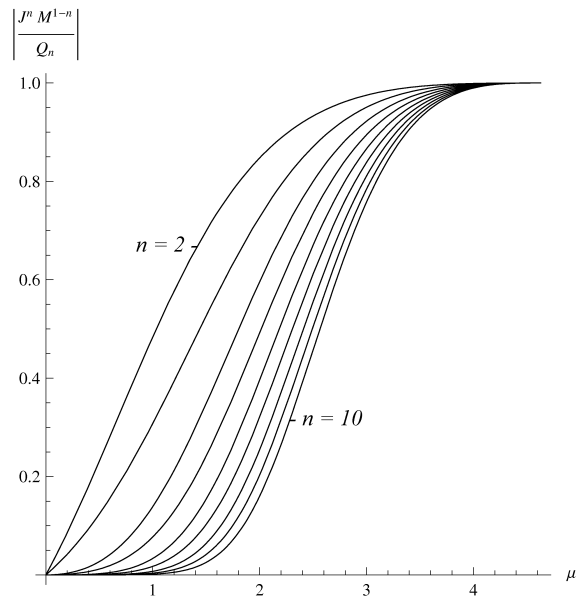}
\caption{\label{fig:disk} The rigidly rotating disk of dust: 
  The conjecture has been verified up to $n=10$.
  The solution depends on a parameter 
  $\mu\,\, (0<\mu<\mu_0 = 4.62966\ldots$), where $\mu \ll 1$
  corresponds to the Newtonian limit and $\mu\rightarrow\mu_0$
  leads to the black hole limit, see \cite{NM93},
  \cite{Meinel02}.}
\end{center}
\end{figure}

The rigidly rotating disk of dust was solved analytically by
Neugebauer and Meinel \cite{NM93,NM95}. The associated multipole
moments were derived shortly thereafter by Kleinw\"achter et al. \cite{KMN95}, see also \cite{Meineletal08}.
Figure \ref{fig:disk} shows that in this case the conjecture is true for $n=2\dots10$.
Moreover, it is interesting that we have $A_{n}\geq A_{n+1}$.

\section{Conclusions and remarks}

Since it is easy to construct solutions of the vacuum Einstein-equations
violating the conjecture, e.g. with the algorithm presented by Manko and Ruiz
in \cite{MR98}, it will be interesting to investigate, 
which requirements on the sources are necessary
to ensure $A_{n}\leq1$ and which are not necessary.

\medskip
We thank Prof. Reinhard Meinel and Dr. David Petroff for inspiring and helpful discussions.

\providecommand{\othercit}{}
\providecommand{\jr}[1]{#1}


\begin{thebibliography}{[10]}

\bibitem{BW71}
 \textsc{J.\,M. Bardeen} and  \textsc{R.\,V. Wagoner}
 ,
 \jr{Astrophys.\ J.} \textbf{167}, 359 (1971).

\bibitem{BF08}
 \textsc{M.~{Bradley}} and  \textsc{G.~{Fodor}}
 ,
 \jr{arXiv:0812.1283} (2008).

\othercit
\bibitem{Filter08}
 \textsc{R.~Filter},
 Multipolmomente axialsymmetrisch station\"arer {R}aumzeiten 
 und die {Q}uadrupol-{V}ermutung,
 diploma thesis, Friedrich Schiller Universit\"at Jena, 2008.

\bibitem{FHP89}
 \textsc{G.~Fodor},  \textsc{C.~Hoenselaers},  and
 \textsc{Z.~Perj\'es}
 ,
 \jr{J.\ Math.\ Phys.} \textbf{30}, 2252 (1989).

\bibitem{Geroch70}
 \textsc{R.~Geroch}
 ,
 \jr{J.\ Math.\ Phys.} \textbf{11}, 2580 (1970).

\bibitem{Hansen73}
 \textsc{R.\,O. Hansen}
 ,
 \jr{J.\ Math.\ Phys.} \textbf{15}, 46 (1973).

\bibitem{Meinel06}
 \textsc{R.~Meinel}
 ,
 \jr{Class.\ Quantum Grav.} \textbf{23}, 1359 (2006).

\bibitem{AKM02}
 \textsc{M.~Ansorg}, \textsc{A.~Kleinw{\"a}chter}, and
 \textsc{R.~Meinel}
 ,
 \jr{Astron.\ Astrophys.} \textbf{381}, L49 (2002).

\bibitem{AKM03b}
 \textsc{M.~Ansorg},  \textsc{A.~Kleinw{\"a}chter},  and
  \textsc{R.~Meinel}
  ,
 \jr{Astron.\ Astrophys.} \textbf{405}, 711 (2003).

\bibitem{Chodosetal74}
 \textsc{A.~Chodos}, \textsc{R.\,L. Jaffe},
 \textsc{K.~Johnson},
 \textsc{C.\,B.~Thorn}, and \\
 \textsc{V.\,F.~Weisskopf}
 ,
 \jr{Phys. Rev.} \textbf{9}, 3471 (1974).

\bibitem{Stoner32}
 \textsc{E.~Stoner}
  ,
 \jr{Mon. Not. R. Astron. Soc.} \textbf{92}, 651 (1932).

\bibitem{NM93}
 \textsc{G.~Neugebauer} and  \textsc{R.~Meinel}
 ,
 \jr{Astrophys.\ J.} \textbf{414}, L97 (1993).

\bibitem{NM95}
 \textsc{G.~{Neugebauer}} and  \textsc{R.~{Meinel}}
 ,
 \jr{Phys.\ Rev.\ Lett.} \textbf{75}, 3046 (1995).

\bibitem{KMN95}
 \textsc{A.~Kleinw\"achter}, \textsc{R.~Meinel}, and
 \textsc{G.~Neugebauer}
 ,
 \jr{Phys.\ Lett. A} \textbf{200}, 82 (1995).

\othercit
\bibitem{Meineletal08}
 \textsc{R.~Meinel}, \textsc{M.~Ansorg},  
 \textsc{A.~Kleinw\"achter},
 \textsc{G.~Neugebauer}, and \textsc{D.~Petroff},
 Relativistic Figures of Equilibrium 
 (Cambridge University Press, Cambridge, 2008).

\bibitem{Meinel02}
 \textsc{R.~Meinel}
 ,
 \jr{Ann. Phys. (Leipzig)} \textbf{11}, 509 (2002).

\bibitem{MR98}
 \textsc{V.\,S. Manko} and  \textsc{E.~Ruiz}
 ,
 \jr{Class. Quantum Grav.} \textbf{15}, 2007 (1998).

\end{thebibliography}
\end{document}